# Perceived Advantage in Perspective: Application of Integrated Choice and Latent Variable Model to Capture Electric Vehicles' Perceived Advantage from Consumers' Perspective

Milad Ghasri, Ali Ardeshiri, Taha Rashidi*

*Research Centre for Integrated Transport and Innovation (rCITI), School of Civil & Environmental Engineering, University of New South Wales, Australia*
*\* Corresponding Author*

**Abstract**
Relative advantage, or the degree to which a new technology is perceived to be better over the existing technology it supersedes, has a significant impact on individuals' decision of adopting to the new technology. This paper investigates the impact of electric vehicles' *perceived* advantage over the conventional internal combustion engine vehicles, from consumers' perspective, on their decision to select electric vehicles. Data is obtained from a stated preference survey from 1176 residents in New South Wales, Australia. The collected data is used to estimate an integrated choice and latent variable model of electric vehicle choice, which incorporates the *perceived* advantage of electric vehicles in the form of latent variables in the utility function. The *design* of the electric vehicle, impact on the *environment*, and *safety* are three identified advantages from consumers' point of view. The model is used to simulate the effectiveness of various policies to promote electric vehicles on different cohorts. Rebate on the purchase price is found to be the most effective strategy to promote electric vehicles adoption.

# 1. Introduction

The history of using electric motors for propulsion in electric vehicles (EVs) starts with the history of batteries in early 18th century (Høyer, 2008). During the last two centuries, EV's market has experienced many ups and downs with an intensive period of technological development and deployment in early 1920s (Høyer, 2008, Westbrook, 2001). The ups and downs in EV's market is mainly determined by the technological improvements in EVs and internal combustion engine (ICE) vehicles, as EVs' competitor alternatives, and the social and political prevailing concerns. Following the environmental paradigm-shift from "*local pollution and noise abatement to global sustainable development and climate change*" (Høyer, 2008) in early 1990s, and the energy crisis in 1970s, which escalated political energy security and non-renewable energy reliance debates (Høyer, 2008), the transport sector, as a major oil consumer and greenhouse gas emitter, has experienced an unprecedented global urge for a transition to alternative fuel vehicles (AFVs), and consequently EVs are back to the spotlight from the beginning of this century (Rezvani et al., 2015). Electrification of transport figures prominently in the European Economic Recovery Plan of 2008 (COM, 2008) and The American Recovery and Reinvestment Act of 2009 (Act, 2009), through grant programs, tax credits, research and development, fleet funding and other measures. In Australia there is no overarching policy for EV at the federal level; however, the majority of state and territory governments have announced EV policy frameworks including The Transition to Zero Emission Vehicles Action Plan in the Capital Territory (ACT, 2018), The Future is Electric in Queensland (Queensland, 2017), and Future Transport Strategy in New South Wales (NSW, 2018).

Considering the political and environmental necessities to replace ICE vehicles with EVs in the transport sector (Sovacool and Hirsh, 2009), a legitimate question to address is the effectiveness of public policies and regulations for promoting EVs. Several groups of stakeholders are involved in AFV adoption including fuel suppliers, car manufactories, consumers, and non-governmental environmental organizations and agencies (Yeh, 2007). The focus of this paper is on the policies and regulations that target consumers. Various behavioural theories, including the theory of planned behaviour (Ajzen, 1991) and the utility maximisation theory (Train, 2009), are used to study the impact of vehicle properties, decision makers' socio-demographic attributes, and consumers' attitudes, habits and values on EV market uptake (Rezvani et al., 2015). High purchase price and operating costs, short driving range, long battery recharge time, and limited recharging stations are often reported as barriers of EVs diffusion process (Hidrue et al., 2011). This paper investigates the main determinants in consumers decision of purchasing EVs with a focus on perceived advantages in the Australian market context, to assist governments with effective resource allocation in promoting EV.

The main contribution of this paper is incorporating consumers' *perception* towards EVs advantage in a discrete choice model based on the utility maximisation theory. Previous studies have used discrete choice specifications to evaluate the likelihood of selecting EV (e.g. see Beggs et al., 1981, Brownstone et al., 2000), or consumers' willingness-to-pay (e.g. see Hidrue et al., 2011, Potoglou and Kanaroglou, 2007, Zhang et al., 2011b). Advanced models are developed to address consumers' preference heterogeneity (Hidrue et al., 2011), and attitudinal impacts (Jensen et al., 2013). However, the implicit impact of consumers' *perception* towards the relative advantages of EVs compared to ICE vehicles, as the superseded alternative, is never brought to the foreground. This is while, social barriers may have an equal impact as technical barriers in hindering the widespread of EVs (Egbue and Long, 2012, Burgess et al., 2013). The theory of innovation diffusion asserts that the perceived relative advantages of an innovation from potential adopters' perspective play a significant role in the innovation diffusion process (Rogers, 2010). In the context of EV, the advantages (and disadvantages) are usually discussed in terms of instrumental factors such as purchase price, operating price, fuel efficiency, etc. We believe, in addition to vehicles attributes, there exist a more enduring structure of individuals' perception towards EV that predisposes their decision. Prior studies show that consumers attitude (Jensen et al., 2013), the provoked emotions from using EV (Moons and De Pelsmacker, 2012), environmental believes and consumer awareness of environmental issues (Lane and Potter, 2007) can affect EVs' adoption rate; however consumers perception of EVs' relative advantage is an overlooked aspect in EVs adoption process.

This study pustulates that consumers' *perception* towards EV advantages has a significant impact on their adoption decision. Moreover, socio-demographic attributes are hypothesised to exert some effects on consumers' *perception*. To investigate this hypothesis, an integrated choice and latent variable (ICLV) model is developed (Ben-Akiva et al., 2002, Vij and Walker, 2016). Parameters of the model are calibrated using an online survey conducted in New South Wales, Australia. The developed model is used to simulate decision makers' reaction towards various supporting policies. To distinguish consumers' dissimilar *perceptions* and its impact on their decision, three sets of socio-demographic attributes are defined to represent Generation X, Generation Y and Generation Z.

The rest of the paper is organised as follows. Section 2 provides a review over studies that have modelled consumers' preference towards EV based on the utility maximization theory. In section 3, the methodology of this study is presented and the details of the developed ICLV model are discussed. Section 4 explains the data collection process and provides some descriptive statistics. Thereafter, the model estimation results are presented in section 5 and the findings are compared with prior studies. The developed model is then used to simulate the sensitivity of various target groups towards supporting schemes and the results are reported in section 6. Finally, the highlights of the study are summarised in the conclusion section.

## 2. Literature review

Most previous demand studies for EVs, that are based on the utility maximisation theory, use stated preferences (SP) data to estimate the model (Sierzchula et al., 2014, Potoglou and Kanaroglou, 2007, Hidrue et al., 2011). Studies that use revealed preference (RP) data, either conduct their analyses on an aggregate level (e.g. see Sierzchula et al., 2014, Yeh, 2007), or enrich the available RP data by SP surveys (e.g. see Brownstone et al., 2000, Axsen et al., 2009). This is primarily due to the low market share of EVs, which makes it difficult to collect an unbiased dataset with an acceptable portion of EV owners (Zhang et al., 2011b), and the multicollinearity between vehicle attributes in revealed preference datasets (Brownstone et al., 2000, Axsen et al., 2009). Moreover, SP surveys enable modellers to study consumers' response to policies and regulations before implementation (Caulfield et al., 2010, Zhang et al., 2011b). In SP surveys, each product is introduced to respondents by a set of its underlying attributes that are expected to be central to respondents' choice (Beggs et al., 1981). The estimated model can be used for calculating EVs' market share (Brownstone et al., 2000, Ito et al., 2013) , calculating willingness-to-pay for various features (Hidrue et al., 2011, Potoglou and Kanaroglou, 2007), or developing agent-based models to simulate market penetration (Zhang et al., 2011a). The importance of taste variation was acknowledged from early studies on this topic, and various methods were utilised to capture consumers' preference heterogeneity. For instance, Dagsvik et al. (2002) estimated different logit models to model the preferences of different age groups and genders, and Hidrue et al. (2011) estimated a latent class model to differentiate EV-oriented and gasoline vehicle-oriented individuals. Jensen et al. (2013) developed a hybrid choice model to capture changes in consumers preferences before and after a trial experience with EV.

A common finding in all these studies is the low preference towards EVs and AFVs in general (Hidrue et al., 2011, Rezvani et al., 2015). The main barriers are reported to be short range, long recharge time, high purchase price and operating costs and insufficient infrastructures for charging (Hidrue et al., 2011).

In addition to the instrumental values, several studies have investigated the impacts from more psychological and social variables such as hedonic and symbolic values (Schuitema et al., 2013, Burgess et al., 2013, Skippon and Garwood, 2011), social norms, environmental attitudes (Hidrue et al., 2011), and neighbourhood conditions (Potoglou and Kanaroglou, 2007) on consumers desire to adopt to EVs. For instance, Moons and De Pelsmacker (2012) extended the theory of planned behaviour with emotional reactions and showed affective components are highly relevant to the usage intention of EVs. Schuitema et al. (2013) used structural equation modelling to capture the impact of hedonic and symbolic values on the intention to adopt to EVs and hybrid vehicles. Hidrue et al. (2011) included pollution in their SP survey and calculated consumers' willingness to pay for pollution reduction in EVs. Pollution reduction was found significant in their study, but they concluded that fuel saving has a higher impact compared to the desire to be green or help the environment. Also, they showed pollution reduction has a lower willingness to pay compared to range and recharge time. Potoglou and Kanaroglou (2007) examine the impact of neighbourhood characteristics at the place of residence on households' preferences towards clean vehicles.

Currently there is no study to examine consumers' *perception* towards EV in their adoption decision. The concept of *perceived* advantage in this study is relevant to the definition of Moons and De Pelsmacker (2012) for visceral and reflective emotions, in the sense that both are more enduring constructs in human psyche that predispose their decision of whether to adopt an EV; however, the perceived advantage in this study relates to perceptions *specific to EVs*, whereas Moons and De Pelsmacker (2012) studied emotions towards car driving in general. The hedonic and symbolic attributes that Schuitema et al. (2013) studied are also specific to EVs, but those attributes, as Schuitema et al. (2013) defined them, are affected by instrumental values so their values changes as the features of the available alternatives vary; whereas the perceived advantage in this study is defined to be a reflection of individuals perception towards the concept of EV, which is gradually formed by acquiring knowledge and experience from this technology, and wouldn't change noticeably by a change in the features of available alternatives. Our definition of perceived advantage is close to what the studied perceptions in the study of Egbue and Long (2012), or what Zhang et al. (2011b) refer to as "awareness" in their study. However, none of these studies incorporated the measured perceptions into the choice modelling context. The study by Jensen et al. (2013) includes an attitudinal component, as a latent variable, to the choice model, but they don't distinguish between various aspects of consumers' perception, as it is discussed in this study.

This study postulates that the perceived advantage of EV has a significant impact on consumers' choice. Besides, it is hypothesised that consumers' perception towards EV can be explained using their socio-demographic attributes. To examine this hypothesis, an ICLV model is developed on a SP survey from potential EV customers in NSW, Australia. ICLV has been previously used to examine the attitudinal and habitual (Paulssen et al., 2014, Vij and Walker, 2016) impacts on decision makers' choice. This study uses the same concept to capture the impact of the perceived relative advantage on consumers adoption decision.

## 3. Methodology

This paper seeks to model observed choices by surveying respondents who also record responses to attitudinal questions. It is hypothesised that both choice and attitudinal responses are influenced by latent variables and we seek to model the choice and attitudinal responses together to give more insight into the processes that motivate respondents' behaviour. In its general formulation, the ICLV model framework is comprised of two components, as illustrated in Figure 1. The two components are the multinomial discrete choice model and the latent variable model. Each sub-model consists of a structural and a measurement component. The utility in the discrete choice component may depend on both observed and latent characteristics of the alternatives and the decision makers. Consistent with the random utility maximization theory, utility as a theoretical construct is operationalized by assuming that individuals choose the alternative with the greatest utility. The latent variable part is rather flexible in that it allows for both simultaneous relationships between the latent variables and MIMIC type models where observed exogenous variables influence the latent variables. Such a specification enables the researcher to disentangle the direct and indirect effects of observed as well as latent variables on the alternatives' utilities. The latent variables themselves are assumed to be measured by multiple indicators representing, in our case, the respondents' answers to Likert-scale survey questions.

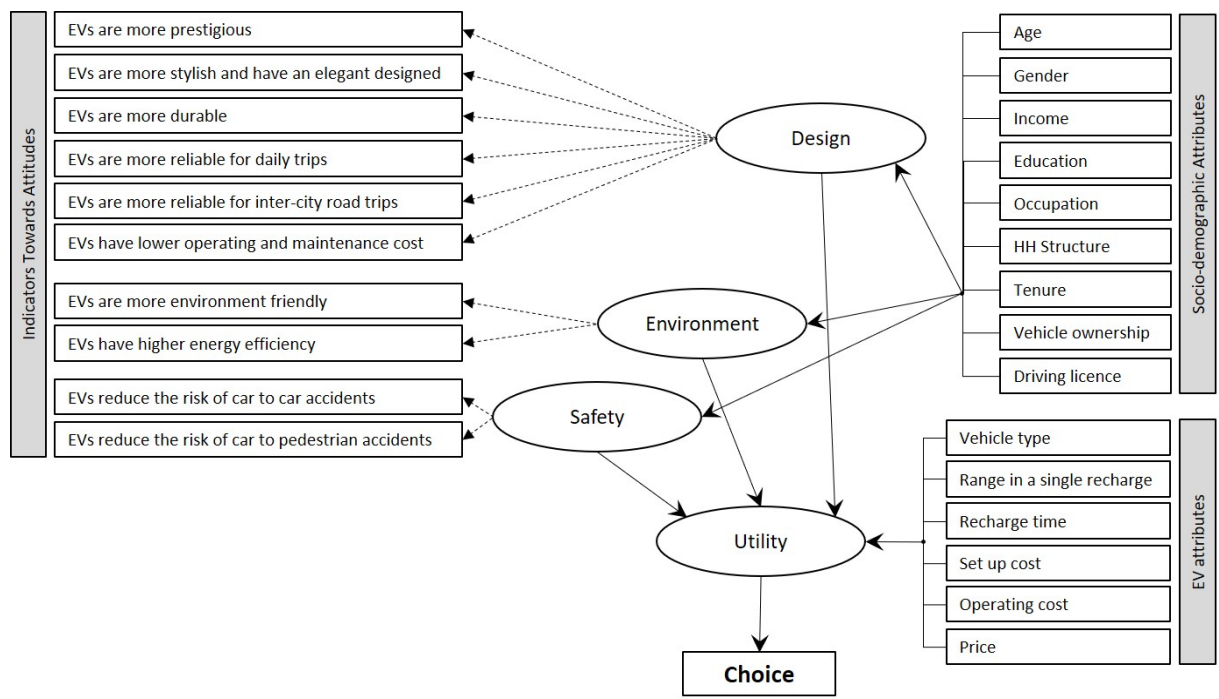

**Figure 1- ICLV flowchart**

### 3.1. The latent variable sub-model

Several studies have shown that an individual's attitude are affected by sociodemographic variables (Paulssen et al., 2014). In this study we examined individuals' attitudes towards the *Design*, *Safety* and *Environmental* effects of EV's. The three latent variables are determined through an extensive investigation in the data and after examining multiple scenarios. We postulate that these personal attitudes influence individual preferences over purchasing an EV. Accordingly, the structural equation for each of the three attitude is defined as shown in equation (1).

$$att_n = Az_n + \delta_n \tag{1}$$

In this equation, $att_n$ denotes the (3 × 1) vector of attitudes for individual $n$. The matrix $A$ donates the unknown regression coefficients, and $\delta_n$ is a (3 × 1) vector of random disturbances assumed to be i.i.d. multivariate normal with mean zero and a (3 × 3) diagonal covariance matrix given by $\sum \delta$ whose non-zero elements are parameters to be estimated.

Model identification requires that the latent variable $att$ be operationalised by multiple manifest indicators $I$. The indicators in this study are responses to a five-point Likert-scale survey questions regarding the level of agreement with statements, as presented in Figure 1. For each of the three latent variables, two or more distinct indicators are used. In all, the measurement model comprises ten indicators. An ordered logit model is used for describing the mapping of indicators onto the latent variables. For example, in constructing the latent variable *Environment*, two indicators, denoted $I_{ENV}$, were used, resulting in the measurement equation (2).

$$I_{ENV,n} = \alpha_{ENV} + \gamma_{ENV} \, att_{ENV,n} + \lambda_{ENV,n} \tag{2}$$

Where $I_{ENV,n}$ is a (3×1) vector donating individual $n$'s responses to the two Likert-scale questions measuring values towards the *environmental* effects of EV's, $\alpha_{ENV}$ is a (3×1) vector of linear regression intercepts, $\gamma_{ENV}$ is a (3×1) vector of loadings and $\lambda_{ENV,n}$ is a (3×1) vector of measurement errors assumed to be i.i.d. multivariate normal across individuals with mean zero and a (3 × 3) diagonal covariance matrix given by $\sum \lambda_{ENV}$ whose non-zero elements are also parameters to be estimated.

For the measurement sub-model to be identifiable, one component of each of $\alpha_{ENV}$ and $\gamma_{ENV}$ has to be fixed so as to set the location and scale[1] for $I_{ENV}$. Measurement equations for each of the remaining two latent variables are similarly formulated.

### 3.2. The discrete choice sub-model

The random utility maximization model is based on assuming the decision-maker (n) faced with a finite set $C_n$ of mutually exclusive alternatives $I$ ($i = 1, ..., I_n$), selects the alternative j which maximises the utility gained. As shown in equation (3), the utility of each alternative ($U_{ni}$) is described as some function of explanatory variables that comprises the systematic (observed) part of the utility function, $V_{ni}$, and some stochastic (unobserved) component, represented by the disturbances $\varepsilon_{ni}$.

$$U_{ni} = V_{ni} + \varepsilon_{ni} \tag{3}$$

---

[1] The scale of the latent variable could alternatively be fixed by constraining the diagonal elements of $\sum \delta$ and $\sum \lambda$, as mentioned in Daziano and Bolduc DAZIANO, R. A. & BOLDUC, D. 2013. Incorporating pro-environmental preferences towards green automobile technologies through a Bayesian hybrid choice model. *Transportmetrica A: Transport Science,* 9**,** 74-106.. However, the two ways are statistically equivalent, and it is usually left to the analyst to choose whichever form is more convenient.

In our case and individual's choice set is composed of three alternatives in total. Two alternatives represent different options for purchasing an EV and alternative three is an option to choose neither of the available options. A linear function of the observable attributes was used in the systematic component of the utility function for any alternative. Regarding the correlation of $\varepsilon_{ni}$ between alternatives, two specification of multinomial logit and nested logit are tested, and MNL specification is found to be more suitable to model this data. The estimated nest parameter in the nested logit specification was not within the meaningful range for this parameter.

Following (Vij and Walker 2016), in this paper we are interested in using the ICLV model to understand the cognitive process underlying decision making, and thus the choice likelihood for the ICLV model has been calculated as choice sub-model interacted with the measurement indicators. Equation (4) shows the systematic part of the utility function in the discrete choice model of this study.

$$V_{ni} = \beta_x(X_n * att_n) + \mu_n \tag{4}$$

In this equation $X_n$ is a matrix of observable attributes regarding the EV property, government support and the market penetration rate and $att_n$ is the matrix of individual attitudes towards *Design*, *Safety* and *Environmental*. All our models were estimated simultaneously using Python Biogeme, an open source freeware designed for the estimation of discrete choice models using maximum simulated likelihood methods (Bierlaire, 2016). For further details regarding the ICLV model we refer the readers to Vij & Walker (2016), Ben-Akiva et al. (2002) and Bolduc et al. (2005).

## 4. Data

The dataset of this study is obtained from an SP survey from a sample of residents in New South Wales, Australia. The survey was administrated online from October 26 to November 7, 2018, through a web-based interface. Respondents were recruited roughly in proportion to the composition of the New South Wales population in terms of key demographic variables, such as age, gender and income. Table 1 shows the descriptive statistics of the sample.

### Table 1 – Descriptive summary

| Variable | Mean (sd.) | Percent |
|---|---|---|
| Age | 45.63 (16.4) | |
| *Age groups* | | |
| between 18 and 30 | 0.23 (0.42) | 23.14% |
| between 31 and 45 | 0.28 (0.45) | 28.07% |
| between 46 and 65 | 0.34 (0.47) | 34.20% |
| between 66 and 85 | 0.15 (0.35) | 14.59% |
| *Gender* | | |
| Female | 0.51 (0.5) | 51.02% |
| Male | 0.49 (0.5) | 48.98% |
| *Income* | | |
| below $52k | 0.29 (0.45) | 29.00% |
| between $53 and $104K | 0.29 (0.46) | 29.28% |
| more than $104k | 0.31 (0.46) | 31.32% |
| not specified | 0.1 (0.31) | 10.41% |
| *Employment* | | |
| Full time | 0.43 (0.49) | 42.75% |
| Part time | 0.19 (0.39) | 19.33% |
| Unemployed | 0.2 (0.4) | 19.98% |
| Retired | 0.18 (0.38) | 17.94% |
| *Education* | | |
| Postgraduate | 0.21 (0.41) | 21.00% |
| Undergraduate | 0.39 (0.49) | 39.41% |
| TAFE Certificate or equivalent | 0.32 (0.47) | 31.88% |
| Other | 0.08 (0.27) | 7.71% |
| *Household Structure* | | |
| Couple with kids | 0.31 (0.46) | 30.76% |
| Couple without kids | 0.36 (0.48) | 35.78% |
| Single parent | 0.05 (0.23) | 5.39% |
| Single | 0.17 (0.38) | 17.38% |
| Other | 0.11 (0.31) | 10.69% |
| *Household Vehicle Ownership* | | |
| No vehicle | 0.04 (0.18) | 3.53% |
| 1 vehicle | 0.5 (0.5) | 49.81% |
| 2+ vehicles | 0.36 (0.48) | 35.50% |
| 3+ vehicles | 0.11 (0.31) | 11.15% |
| *Dwelling type* | | |
| House | 0.67 (0.47) | 67.10% |
| Apartment | 0.22 (0.41) | 21.93% |
| Other | 0.11 (0.31) | 10.97% |
| *Tenure type* | | |
| Owner | 0.36 (0.48) | 36.15% |
| Owner with mortgage | 0.33 (0.47) | 32.53% |
| Renter | 0.29 (0.46) | 29.46% |
| Other | 0.02 (0.14) | 1.86% |

The SP section in this survey is designed to elicit preferences for EVs from presentation of hypothetical scenarios. During the survey, respondents encountered eight SP tasks where two hypothetical scenarios for EVs were presented to them and they could choose either or none. In other words, in each task three alternatives are available to choose from.

There groups of attributes are considered to describe EV alternatives. The first group encompasses vehicle properties such as price and technical features. The second group contains various support schemes from the Government, and the third group includes EVs' market uptake. Table 2 shows the variables and their corresponding levels. The vehicle property category includes body type, purchase price, set up cost, operating cost, recharge time and range in a single recharge. These attributes are selected according to previous studies (Beggs et al., 1981, Brownstone et al., 2000, Caulfield et al., 2010, Sierzchula et al., 2014). Sierzchula et al. (2014) considered cost, range and recharge time as three of the technological barriers in adopting EVs. Beggs et al. (1981) and Brownstone et al. (2000) found body type to be significant in consumers' preference towards EVs.

The support schemes are primarily adopted from previous studies. Sierzchula et al. (2014) divided government financial incentives into technology specific policies, which include subsidies to EV consumers, and technology neutral policies, such as emission tax. Potoglou and Kanaroglou (2007) examined willingness-to-pay for access to HOV lanes, parking rebate and tax deduction. The support scheme category includes infrastructure improvement such as establishing fast charge stations or providing access to bus lanes for EVs, and financial incentives such as rebates on purchase cost or parking fees, or discounts on stamp duty or energy bill.

The last section pertains to market uptake. The portion of EV registration can communicate a confirming signal from peers on EV performance. Higher portions of EV registration suggests that higher number of drivers have trusted this new technology and this is an encouraging signal for those who are awaiting peers confirmation before adopting to a new technology (Rogers, 2010).

**Table 2 – Alternatives' attributes and levels in the SP task**

| Attribute | Min | Max | Number of levels | Level(s) |
|---|---|---|---|---|
| **Vehicle Property** | | | | |
| Vehicle body type | - | - | 6 | Small Hatchback; Small Sedan; Large Sedan; Small SUV; Large SUV; Minivan |
| Price (condition on available budget) | $25,000 | $55,000 | 4 | $25,000; $35,000; $45,000; $55,000 |
| | $55,000 | $100,000 | 4 | $55,000; $70,000; $85,000; $100,000 |
| | $100,000 | $160,000 | 4 | $100,000; $120,000; $140,000; $160,000 |
| Set up cost | $1,000 | $3,250 | 4 | $1,000; $1,750; $2,500; $3,250 |
| Operating cost | 3 c/km | 12 c/km | 4 | 3 c/km; 6 c/km; 9 c/km; 12 c/km |
| Recharge time | 0.5 hr | 7.5 hr | 8 | 0.5 hr; 1.5 hr; 2.5 hr; 3.5 hr; 4.5 hr; 5.5 hr; 6.5 hr; 7.5 hr; 8.5 hr |
| Range in a single recharge | 120 km | 540 km | 8 | 120 km; 180 km; 240 km; 300 km; 360 km; 420 km; 480 km; 540 km |
| **Support Scheme** | | | | |
| Distance between fast charge station | 5 km | 20 km | 4 | 5 km; 10 km; 15 km; 20 km |
| Access to bus lane | | | 2 | Yes; No |
| Rebate on upfront cost | $3,000 | $10,000 | 4 | NA; $3,000; $6,500; $10,000 |
| Rebate on parking fee until 2025 | $100 | $400 | 4 | NA; $100; $250; $400 |
| Energy bill discount until 2025 | 25% | 100% | 4 | NA; 25%; 75%; 100% |
| Stamp duty discount | 5% | 25% | 4 | NA; 5%; 15%; 25% |
| **Market penetration** | | | | |
| Current portion of EVs sold | 1 out of every 100 | 90 out of every 100 | 4 | 1; 30; 60; 90 |

## 4.1. Experimental design

The attribute level values used in specific choice tasks were defined by an efficient experimental design generated using NGENE software. We generated a design using the D-efficiency (Ardeshiri and Rose, 2018, Scarpa and Rose, 2008). D-efficiency design strategies produce significantly improved results, in a statistical sense of relative efficiency, then the more traditional orthogonal design (Rose, Bliemer, Hensher, & Collins, 2008).

The final design had a D-error of 0.0092 and included 144 choice tasks in 18 blocks, providing each participant with 8 repeated choice scenarios. Each individual was given 8 hypothetical tasks to complete and was urged to treat each task independently of the others. In each task participants had the option to choose between two electric vehicles. The no option was available as the third alternative. Participants were also reminded to keep in mind their budget range they had anticipated for purchasing a vehicle, as well as their household income and all other things that this income is spent on. To ensure that the participants took the survey seriously, a short cheap talk script was developed using guidance from Morrison and Brown (2009). Cheap talk is a technique used in SP surveys to remind participants that they should make choices as if they really had to pay. Cheap talk has been shown to be effective at reducing the potential for hypothetical bias in choice experiments (Ardeshiri et al., 2018, MacDonald et al., 2015, Tonsor and Shupp, 2011, List et al., 2006). Figure 2 presents an example of the DCE task.

**Figure 2: Screen capture of an example of the DCE task**

[Figure: Screenshot of UNSW survey Task 1 of 8 showing an Electric Vehicle choice experiment comparing Option 1 (Large sedan, 420 km range, 4.5 hours recharge, $2,500 setup, 3 cents/km, $55,000, charging every 15 km, bus lane access yes, $3,000 rebate, $100/year parking rebate, 25% energy bill discount, 25% stamp duty discount, 1 out of 100 cars sold) vs Option 2 (Small hatchback, 120 km, 3.5 hours, $1,000, 3 cents/km, $100,000, every 5 km, no bus lane, $6,500, $400/year, N/A, 25%, 1 out of 100), with a third option "I would NOT purchase either of the Electric Vehicle options".]

## 5. Results

Several model specifications with different number of latent variables, in the structural component of the model were tested to obtain the presented specification in this section. This model is the most reliable specification with the highest goodness-of-fit measure and with a large number of meaningful significant coefficients.

### 5.1. Structural part

After examining several specifications, the construct with three latent variables was selected as the most suitable specification. An exploratory factor analysis (EFA) also suggested the same number of latent variables to summarise respondents' perception. Figure 1 shows the relation between latent variables and measurement questions. As shown in this figure, the first latent variable, which we refer to as *design*, encompasses consumers' perception about performance, reliability and aesthetic aspects of EVs. The second latent variable reflects on consumers perception about EVs environmental impacts, and the third latent variable pertains to safety aspects of EVs. Noteworthy to mention that Egbue and Long (2012) also considered technical features, environmental impacts, and safety to study individuals perception towards EV.

Table 3 shows the estimated coefficients for the structural part. The large number of statistically significant variables in this table confirms the hypothesis that socio-demographic attributes influence consumers perception about EVs. Out of the 69 incorporated variables, the estimated coefficient for 56 and 59 variables are significant with 95% and 90% confidence level respectively.

*5.1.1. Individual level attributes*

Age is known to have a non-linear impact; therefore, various formats for this variable, including stepwise, piecewise linear, quadratic and cubic formats, are examined, and the meaningful format with highest goodness-of-fit is selected. To illustrate the impact of age, all other variables are set to zero and the profile of variations in the latent variables against age is plotted in Figure 2. According to this figure, *environment* has the lowest and *safety* has the highest variations. All the curves start with a decreasing trend but after reaching a trough they switch to an increasing trend. The trough for *safety* and *design* occurs between 60 to 70, and the trough for environment at around 50.

The binary variable for females is significant in all three latent variables suggesting that females are inclined towards believing more efficient design, higher safety and lower environmental impacts for EVs. Caulfield et al. (2010) showed that men are less likely to purchase hybrid electric vehicles, and Hidrue et al. (2011) showed men are more likely to be gasoline vehicle-oriented rather than EV-oriented.

Across all the latent variables, the coefficient for postgrad is higher than undergraduate, and they are both higher than certificate. Consistent with prior studies (e.g. see Potoglou and Kanaroglou, 2007) it suggests that higher education levels increase the perceived advantages in EVs, which consequently increases the likelihood of buying an EV. With the same token, compared to part time workers, full time workers consider higher values in EVs.

The estimated coefficients for income also indicate a non-linear impact on perception. In the survey, income is aggregated in three brackets. For identifiability issues, the coefficient for middle-income group (income between $55k and $104k) is set to zero. According to Table 3, people in lower-income category believe EVs are safer vehicles with a more stylish and efficient design. On the other hand, people in higher-income bracket perceive EVs to be more environment friendly.

*5.1.2. Household level attributes*

According to Table 3, vehicle ownership negatively affects peoples' perception towards EVs. As shown in Table 1, more than 96 percent of the respondents have 1 vehicle or more, and nearly 98 percent of vehicle owners have ICE vehicles. Therefore, nearly 94 percent of individuals currently have access to ICE vehicles and according to Table 3 these people do not perceive EVs to be advantageous. The negative impact of vehicle ownership on consumers preference towards EV is reported in previous studies as well (e.g. see Zhang et al., 2011b). This finding is substantially important, because ICE owners are the potential consumers for EVs, but they seem to be satisfied with their current choice and they do not perceive additional advantages in adopting EVs. Rogers (2010) consider the perceived relative advantage of an innovation to be positively related to its rate of adoption; therefore, the relatively lower perceived advantages by ICE owners can be a major barrier to EVs diffusion.

Regarding household structure, being a part of a "couple with kids" household is found to have a positive impact on people's perception towards EV. The coefficients for this variable are significant for all the three latent variables which suggests people with this household type consider EVs to be safer, more environment friendly, and better designed. Considering *design*, the other significant coefficient is for singles. The negative sign of this coefficient indicates that singles do not believe that EVs a superior design compared to conventional vehicles. For *environment*, all the household type coefficients are found positive and significant, suggesting that they all see EVs to be more environment friendly.

Regarding *safety*, the only two significant household types are "couple with kids" and "single parents". This finding suggests that parents believe EVs are safer, and this belief may come from their concerns about their children safety.

**Table 3 – Estimation results for the structural component**

| Variable | Design | Environment | Safety |
|---|---|---|---|
| Constant | 3.97 | 5.41 | 3.05 |
|  | (19.32) | (14.46) | (8.56) |
| *Socio-demographic* | | | |
| Age | -3.33 | -9.36 | 2.58+ |
|  | (-2.54) | (-3.9) | (1.1) |
| Age Squared | -8.71 | 13.6 | -27.5 |
|  | (-3.04) | (2.61) | (-5.38) |
| Age cubed | 12.2 | -5.61+ | 27.4 |
|  | (6.22) | (-1.58) | (7.88) |
| Female | 0.086 | 0.266 | 0.347 |
|  | (5) | (8.45) | (11.14) |
| *Education* | | | |
| TAFE Certificate or equivalent | -0.15 | 0.174 | -0.053+ |
|  | (-4.65) | (2.95) | (-0.92) |
| Postgraduate | 0.444 | 0.355 | 0.736 |
|  | (12.1) | (5.37) | (11.21) |
| Undergraduate | 0.196 | 0.266 | 0.157 |
|  | (5.99) | (4.45) | (2.66) |
| *Employment* | | | |
| Full time | 0.362 | 0.155 | 0.368 |
|  | (15.53) | (3.71) | (8.9) |
| Part time | 0.122 | -0.065+ | 0.354 |
|  | (5.25) | (-1.53) | (8.4) |
| *Household Structure* | | | |
| Couple with kids | 0.49 | 0.431 | 0.665+ |
|  | (16) | (7.88) | (12.21) |
| Couple without kids | -0.042+ | 0.283 | -0.041 |
|  | (-1.39) | (5.06) | (-0.73) |
| Single parent | 0.053+ | 0.209 | 0.449 |
|  | (1.21) | (2.66) | (5.73) |
| Single | -0.128 | 0.363 | -0.019+ |
|  | (-3.83) | (5.85) | (-0.32) |
| *Vehicle ownership* | | | |
| 1 Vehicle | -0.093 | -0.343 | -0.635 |
|  | (-2.08) | (-4.11) | (-7.99) |
| 2 Vehicles | -0.207 | -0.379 | -0.954 |
|  | (-4.31) | (-4.27) | (-11.18) |
| 3+ Vehicles | -0.317 | -0.217 | -1.34 |
|  | (-5.99) | (-2.22) | (-14.16) |
| *Income* | | | |
| Below $54k | 0.409 | 0.017+ | 0.396 |
|  | (18.5) | (0.44) | (10.35) |
| Above $104k | 0.038+ | 0.111 | 0.067+ |
|  | (1.87) | (2.9) | (1.79) |
| *Accommodation* | | | |
| House | -0.047+ | -0.161 | 0.123 |
|  | (-1.71) | (-3.17) | (2.47) |
| Apartment | -0.206 | -0.179 | -0.05+ |
|  | (-6.85) | (-3.2) | (-0.93) |
| Owner | 0.524 | -0.581 | 0.436 |
|  | (8.99) | (-5.09) | (4) |
| Owner with mortgage | 0.349 | -0.585 | 0.269 |
|  | (5.98) | (-5.1) | (2.45) |
| Renter | 0.379 | -0.568 | 0.205+ |
|  | (6.5) | (-4.96) | (1.88) |

+ Not significant with 95% confidence level

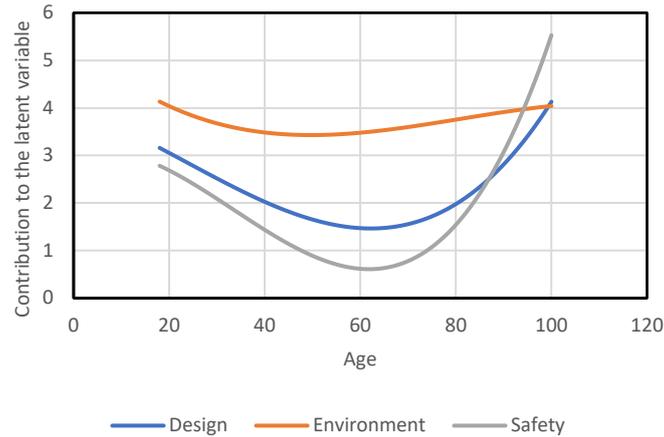

**Figure 2- Profile of age contribution to the latent variable**

### 5.2. Discrete choice component

The estimated coefficients for the discrete choice component ($\beta$ in equation 4) are presented in Table 4. The incorporated variables in the utility functions are divided into vehicle properties, supporting schemes, market uptake and interacted variables with the latent variables.

*5.2.1. Vehicle property*

Corresponding to the six levels of body type (as specified in Table 2), six binary variables are defined and due to identifiability issues the binary variable for minivan is set to zero. The positive coefficient for hatchback, small sedan and small SUV indicates that people prefer these body types compared to minivan. As expected and confirmed in previous studies (e.g. see Beggs et al., 1981, Brownstone et al., 2000, Caulfield et al., 2010), all the variables related to cost have a negative coefficient in the utility function. As expected for a rational consumer, higher purchase price, higher set up cost or higher operating costs will reduce the likelihood of choosing EVs. The coefficient for recharge time is negative indicating that individuals prefer vehicles with lower recharge times. Beggs et al. (1981) pinpointed the relative low range and high recharging time of EV (at the time of their study) to be two main barriers in its market uptake.

*5.2.2. Support scheme*

Six support schemes to promote EVs are examined and the two schemes of "rebates on upfront costs" and "discount on energy bills until 2025" are found significant in individuals' decision to purchase EV. Rebate on upfront costs is a one-off financial incentive to reduce purchase price for consumers and energy bill discount until 2025 is an ongoing discount on operating costs. Prior studies also reported that reducing monetary costs and purchase tax relieves are effective policies in the adoption process of clean vehicles and AFV (e.g. see Caulfield et al., 2010, Potoglou and Kanaroglou, 2007).

Infrastructure incentives of "access to bus lanes" and "distance between fast charge stations", or financial incentives on "stamp duty discount" and "rebate on parking fees until 2025" are not found significant in this study. After conducting the likelihood test, they are excluded from the utility function. This can be because individuals do not see much value in these support schemes and their influence is so diminutive that cannot be captured in the dataset of this study. There are evidences in the literature suggesting that rebate on parking fees or access to high occupancy vehicle lanes are not significant (Potoglou and Kanaroglou, 2007). Regarding the impact of charging stations, literature have contradicting findings. While Brownstone et al. (2000) showed that availability of charging station is a predictor (in a SP data), Sierzchula et al. (2014) didn't find a significant impact from charging stations, when analysing the market uptake of EV and plugged in hybrid vehicles in 30 countries,.

Table 4 – Estimation results for the discrete choice component

| Variable | Estimate |
|---|---|
| ***Vehicle property*** | |
| Hatchback | 0.49 (0.058) |
| Small Sedan | 0.417 (0.058) |
| Small SUV | 0.499 (0.057) |
| Purchase price | -7.98 (0.283) |
| Set up cost | -0.038 (0.02) |
| Operating costs | -0.46 (0.049) |
| Recharge time | -0.477 (0.073) |
| ***Supporting scheme*** | |
| Rebate on upfront cost | 0.183 (0.046) |
| Energy bill discount until 2025 | 0.309 (0.059) |
| ***Market uptake*** | |
| Portion of EVs sold | 0.344 (0.053) |
| ***Interacted latent variables*** | |
| Design * Purchase price | 2.38 (0.094) |
| Environment * Range in a single recharge | 0.023 (0.004) |
| Safety * Large SUV | 0.139 (0.025) |
| Safety * Large Sedan | 0.084 (0.024) |
| ***Goodness of fit measures*** | |
| Log likelihood | -118279.273 |
| Rho square | 0.373 |

### 5.2.3. Market uptake

Current market uptake is known to be an influential factor in the diffusion of a new product or idea (Rogers, 2010). On one hand, conformity encourages people to adopt behaviours and lifestyles that are in accordance with social accepted conventions and their peers' common practice. On the other hand, peers' adopting a new technology send an approval signal from peers (Rogers, 2010). This would increase the desire to adopt the new technology as people find their peers' assessment more reliable compared to experts' opinion broadcasted in mass media (Rogers, 2010). The findings of this study confirm this hypothesis, as the estimated coefficient for "portion of EVs sold" is positive.

### 5.2.4. Interacted latent variables

As mentioned in section 3.2, the latent variables in the utility function are interacted with vehicle attributes. Interacting vehicle attributes with socio-demographic attributes is common in the literature (Beggs et al., 1981, Brownstone et al., 2000). In this study, the reflection of socio-demographic attributes in individuals' perception is interacted with vehicle attributes. As shown in Table 4, four interacted cases are found significant. The first case shows the combined impact of purchase price and *design*. As discussed in section 5.2.1, purchase price is found negative in the likelihood function so higher purchase prices reduces the probability of purchasing EV. However, when this variable is interacted with *design*, its coefficient becomes positive, which means those who perceive EVs to be more advantageous regarding its design, are willing to pay more to purchase it.

Range in a single recharge is one of the technology related properties of EV that determines how often the vehicle needs recharging. Dagsvik et al. (2002) showed range has a positive impact in selecting AFV. Brownstone et al. (2000) considered the impact of range to be non-linear and showed respondents value an increase in range more highly when starting from a lower base. Accordingly, a positive coefficient for this variable is expected, as higher values for this variable indicates lower frequency of recharging. When this variable is interacted with *environment*, it suggests that those who believe EVs are more environment friendly are more sensitive towards range. Longer ranges are related to higher efficiency and lower environmental impacts, so those who see EVs to be more environment friendly are more sensitive towards range.

*Safety* is interacted with body types of "large sedan" and "large SUV" and both variables have a positive coefficient in the utility function. This observation indicates that people who believe EVs are safer have a higher tendency towards large body types such as large sedan and large SUV. Larger vehicle body can be considered as a stronger shield which provide a safer place for driver and passengers.

## 6. Discussion

The complex structure of the ICLV model, especially with the interacted latent variables in the utility function of the discrete choice component, makes it difficult to evaluate the impact of independent variables from the estimated coefficients. In this section, the developed ICLV model is used to simulate individuals' responses to various scenarios, in order to investigate the effectiveness of several support schemes. In this exercise, vehicle attributes are defined based on the Tesla Model S. For socio-demographic attributes, three generations of Generation X, Y and Z are selected in this practice.

### 6.1. Vehicle properties

The Tesla Model S 2019 is selected as a commercially available EV in the market (TESLA, 2019). This vehicle is a large sedan and the purchase price for it is nearly $100,000 AUD. There is an additional cost of $10,000 AUD for local registration costs, luxury car tax, local stamp duty and delivery fee which are considered as the set-up cost (all the figures are for New South Wales). The standard range for this vehicle is reported as 450 km, the recharge time for this vehicle is claimed to be 75 minutes (TESLA, 2019), and the operating costs for this vehicle in Sydney is estimated to be 6.12 cent per kilometre.

### 6.2. Consumers target groups

As discussed in section 5, socio-demographic attributes exert some effect on consumers' perception towards EV. For the simulation practice of this paper, we consider six different set of socio-demographic variables, as male and female representatives for Generation X, Y and Z.

*6.2.1. Generation X*

Generation X or Gen X refers to the demographic cohort with the birth years between the early 1960s and the early 1980s. Gen X includes experienced drivers with the age between 40 to 60 (at the time of this study), who have witnessed the recent evolutions in the automobile industry and probably have trade more vehicles compared to drivers from generations Y and Z. We consider 50 as the representative age for Gen X. For other socio-demographic attributes, we used the most frequent observed values for

40 to 60 year-old respondents (Gen X age range) in the dataset of this study. Since gender is shown to have a significant impact on EV preference, we distinguished between male and females for Gen X, as well as, the other two generations.

*6.2.2. Generation Y*

Generation Y or Gen Y, also known as Millennials, are the demographic cohort following Gen X and preceding Generation Z. Typically, the early 1980s is referred to as the starting birth years and mid-1990s as the ending birth years. The age range for Gen Y varies between 25 to 40 (at the time of this study) and we used 37 as the representative age for Gen Y. Similar to Gen X, we distinguish between male and female, and the rest of socio-demographic attributes are set to their mode values for Gen Y in the dataset of this study.

*6.2.3. Generation Z*

Generation Z or Gen Z is the cohort after the Millennials who are born between mid-1990s and mid-2000s. Gen Z includes young drivers who are either contemplating purchasing their first vehicle or have recently gone through this experience. We consider 20 as the representative age for Gen Z, and for other socio-demographic attributes, we used their mode value for Gen Z in the dataset of this study.

Table 5 shows the utilised socio-demographic attributes in the simulation practice. According to this table, Gen X and Gen Y have more attributes in common. The highest education level for Gen X and Gen Y is undergraduate, but for Gen Z it is "TAFE certificate or equivalent". The employment status for Gen X and Gen Y is full time, whereas respondents in Gen Z are mostly part time employees. Household structure and income are also the same for Gen X and Gen Y, but they differ from the values for Gen Z. While all the hypothetical representatives of the three generations live in a house, they have different tenure status, with Gen Z being renter, Gen X being owner and Gen Y being owner with mortgage.

**Table 5 – Socio-demographic attributes of the target groups**

| Variable | Gen Z | Gen Y | Gen X |
|---|---|---|---|
| Age | 20 | 37 | 50 |
| Education | TAFE Certificate or equivalent | Undergraduate | Undergraduate |
| Employment | Part time | Full time | Full time |
| Gender | Male/Female | Male/Female | Male/Female |
| Household structure | Other | Couple with kids | Couple with kids |
| Household vehicle ownership | 1 | 1 | 1 |
| Income | Between $53 and $104k | More than $104k | More than $104k |
| Dwelling type | House | House | House |
| Tenure status | Renter | Owner with mortgage | Owner |

### 6.3. Perceived advantages and disadvantages

The utilised model of this study enforces the calculation of perceived advantages before calculating the probability of selecting EV. To simulate the probability of selecting EV, we first need to simulate the value of latent variables for each cohort. Table 6 displays the calculated latent variables for each cohort. According to this table, women see higher values in EVs across the three generations, because their latent variables are higher compered to men. This can explain the reported higher preference is women towards EVs (e.g. see Caulfield et al., 2010). With the same token, people from Gen Y perceive EVs to have better design and to be more environment friendly. When it comes to *safety*, people from Gen Z see more advantages in EVs.

Table 6 – Calculated latent variables

| Variable | Gen Z | | Gen Y | | Gen X | |
|---|---|---|---|---|---|---|
| | Male | Female | Male | Female | Male | Female |
| Design | 3.26 | 3.35 | 3.46 | 3.54 | 3.12 | 3.21 |
| Environment | 3.07 | 3.34 | 3.40 | 3.66 | 3.31 | 3.57 |
| Safety | 2.68 | 3.03 | 2.64 | 2.99 | 2.07 | 2.42 |

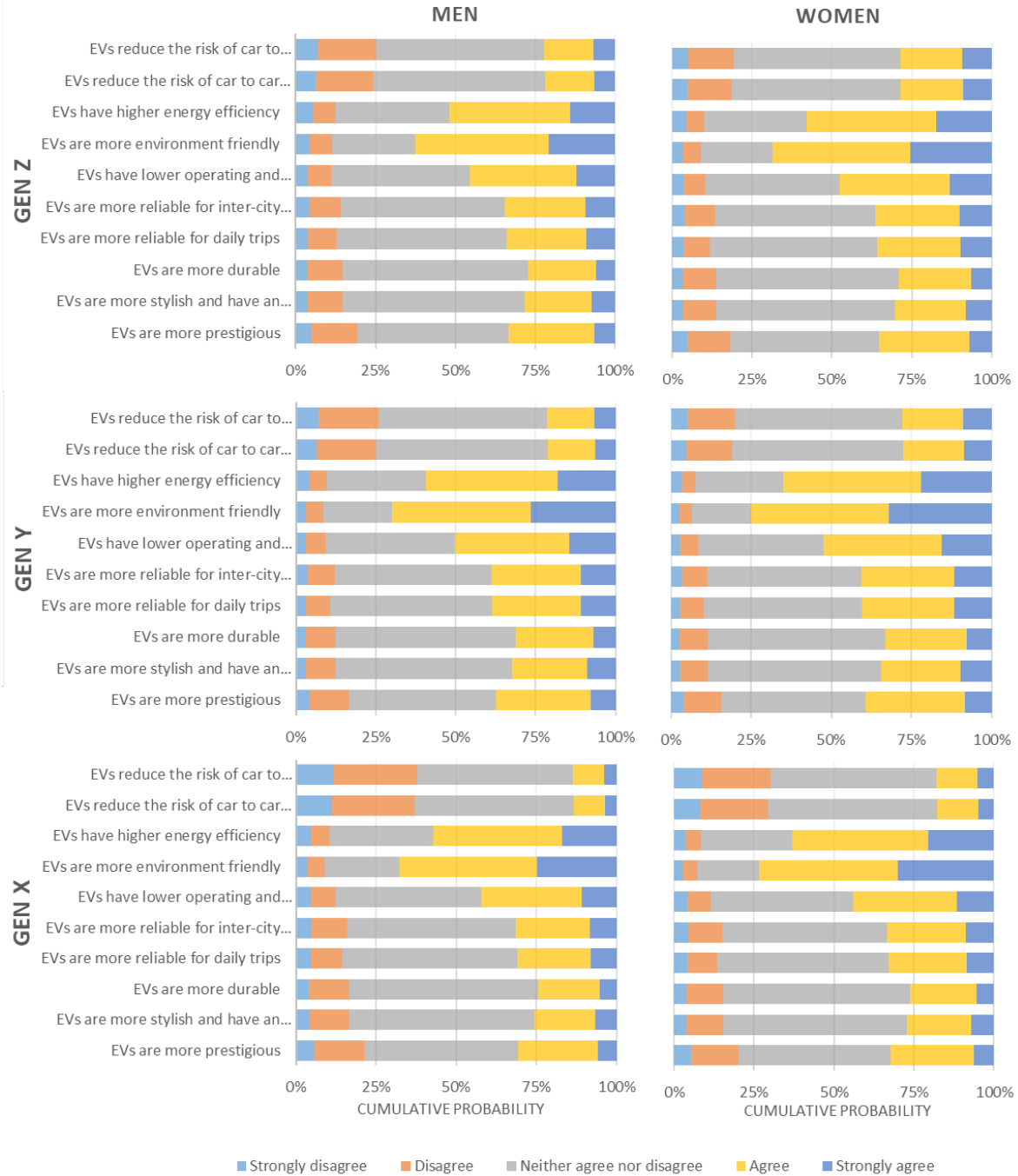

**Figure 3- Simulation results for the measurement component**

After calculating the latent variables, the structural component of the model can be simulated. The estimated results for the ordered logit models, which captures the probability of selecting each level in the Likert scale questions, are precluded for brevity. However, the simulated outputs of the ordered logit models are presented in Figure 2. The six plots in Figure 2 correspond to the gender specific representatives of the three generations. The bars in each plot correspond to the ten Likert scale questions, and the horizontal axis shows the cumulative probability for the five levels of "strongly disagree", "disagree", "neither agree nor disagree", "agree" and "strongly agree". According to this figure, women have a higher agreement rate (strongly agree and agree levels) across all the questions and in all the generations. Gen X has higher disagreement (strongly disagree and disagree) rate compared to Gen Y and Gen Z. This is more noticeable for the first two questions which pertain to safety. The other distinct pattern in these plots is the high agreement rate for the second two questions which are related to higher energy efficiency for EVs and lower environmental impact.

### 6.4. Policy assessment

Government policies on demand side are essential in EV market, not only because consumers' subsidies are found to have a significant effect on EV market uptake, but mainly due to the *market failure* condition in the market of EVs and other eco-friendly innovations (Sierzchula et al., 2014). In economics, *market failure* refers to a situation where the allocation of goods or services by a free market is not efficient. For EVs, a primary advantage is the reduction in emitted pollution from which the entire population will benefit, but it is impossible for private automobile industry to charge all the members of public for this benefit (Sierzchula et al., 2014). Therefore, government intervention in EV market is essential.

In the literature, the most effective policies are reported to be financial incentives (Sierzchula et al., 2014). Besides, one-off incentives on the upfront costs are found to be more effective compared to ongoing incentives on maintenance costs. In this paper, we analyse the impact of one-off incentives as subsidies to manufactories to lower the purchase price (scenario 1), or direct payment to consumers as rebate on upfront cost (scenario 2). Technological improvements are represented by increasing the range (scenario 3) and reducing the recharge time (scenario 4). As an ongoing financial incentive, discount on electricity bill is studied (scenario 5), as electricity costs is the main determinant in EVs' operating costs (Dijk et al., 2013, Sierzchula et al., 2014). To put the results in perspective, the impact of market uptake on individuals' preference towards EV is also analysed (scenario 6). Figure 3 summarises the simulation results for the six scenarios.

Each plot in Figure 3 contains six lines corresponding to the six subject cohorts. The vertical axis shows the probability of selecting EV and the horizontal axis shows the range or variations for the variable of interest in each scenario. The estimated probabilities in this practice are somewhat optimistic compared to the current market share of EV. This issue is common in almost all SP studies (Axsen et al., 2009). This discrepancy is usually justified by missing important factors, such as supply constraints, from the SP survey, or hypothetical or social desirability bias (Axsen et al., 2009). In the context of EV market uptake, the phenomenon of "attitude-action gap" can explain this discrepancy to some extent. This phenomenon indicates that despite the high level of concern declared with environmental issues, when it comes to purchasing a new car, individuals consider little importance on environmental impacts (Lane and Potter, 2007, Sierzchula et al., 2014). In an SP survey, respondents' environmental attitudes play a significant role which lead them to select more environment friendly options, but environmental concerns have little relation to individuals action (Lane and Potter, 2007).

A general observation across all the six scenarios is that women have higher probability of selecting EV compared to men. Consistent with prior studies, this observation stem from the fact that women perceive higher benefits in EVs (refer to section 6.3). Also, probabilities for Gen Y is always higher than probabilities for Gen Z and they are both higher than probabilities for Gen X. Although in the simulation practice everyone faces the same choice set up where the vehicle attributes are the same, the choice outcome varies due to the dissimilar perceptions. As discussed in section 6.3, the representatives of the three generations perceive EV differently, and their preference to purchase an EV varies accordingly.

### *6.4.1. One-off incentivisation*
Scenario 1 represents subsidy to manufactories in order to reduce purchase price, and scenario 2 represent subsidy to end users as rebate on the upfront cost. Comparing the simulation results for these two scenarios suggests that providing financial incentives to the end users is more effective compared to providing the same incentives to the producers. The amount of incentive varies from 0 to 50 thousand dollars in both scenarios. In scenario 1, when the incentive is allocated to manufactories, it does not show significant impact on Gen Y. The curves corresponding to men and women for Gen Y have almost a flat trend around 50 percent and 55 percent respectively. For Gen Z the probability of selective EV raises from 36.9 to 46.4 percent for men, and from 43.2 to 50.4 percent for women. Gen X are the most sensitive generation towards this incentivisation with an increase from 28.9 to 41.60 percent for men, and from 34.6 to 45.5 percent for women. Shifting the attention to scenario 2, when the same amount of incentive is offered to the end users as a rebate on upfront cost, it shows higher impacts on people's preference towards EV. In scenario 2, the impact is not much different across generations. Providing a rebate on upfront cost of 50 thousand dollars increases the probability of selecting an EV by approximately 22 percent.

The effectiveness of financial incentives to consumers is already acknowledged in the literature (Sierzchula et al., 2014, Krupa et al., 2014); however, the methodology of this study reveals new aspects of this phenomenon. The mechanism of providing financial incentives plays a role in its effectiveness. The reason behind dissimilar impacts of identical incentives stem from individuals' perception. When 50 thousand dollars incentive is assigned to manufactories, the purchase price of the EV will drop from 100 to 50 thousand dollars, but when the same incentive is offered to consumers, the nominal price of the vehicle is still 100 thousand dollars. This behaviour can be considered as an example from nudge theory (Thaler and Sunstein, 2009), where decision makers act instinctively rather than logically. Even though the market price for the vehicle is identical in both scenarios, consumers value the rebate on upfront cost more as their perception is shifted to associate higher values with the vehicle in this case. When the incentive is offered to manufactories to reduce the purchase price, from consumers perspective, the purchased vehicle is worth 50,000 dollars, but when the incentive is offered as rebate on upfront cost, then the vehicle is worth 100,000 dollars from consumers perspective and they only pay for half of it. Therefore, they are more likely to purchase an EV.

Considering the mechanism of capturing this discrepancy in the developed model, purchase price appears twice in the discrete choice component. As shown in Table 4, purchase price is included as one of the vehicle properties with a negative coefficient, and it is interacted with *design* and the estimated coefficient is positive. The negative coefficient for purchase price implies a decreasing impact from this variable on the desire to select EV, whereas the positive coefficient for the interactive term adjusts this decreasing impact according to the perception towards EV's *design*. Those who believe EVs have a superior *design* are more willing to accept higher purchase prices. This is the reason that changes in the purchase price in scenario 1 does not have a noticeable impact on Gen Y. In other words, the perceived advantage in EV's *design* is high enough that it balances the disutility of high purchase prices. On the other hand, the positive coefficient for rebate on upfront cost in Table 4 indicates an increasing impact from rebate on upfront cost on the desire to purchase EV. This impact is roughly the same for everyone regardless of their perception towards EV.

This phenomenon can also be explained by theories related to EV symbolism (Burgess et al., 2013, Skippon and Garwood, 2011), which suggest products such as motor vehicles not only serve consumers with their practical needs such as mobility; but also, they help with creating a social image for their users by sending symbolic messages to the members of society (Miller, 2009, Skippon and Garwood, 2011). Burgess et al. (2013) assert that symbolic benefits sometimes override the rational utility-based calculations. For instance, purchasing an EV may indicate higher concerns and a pro-environmental behaviour with the environment (Rezvani et al., 2015), or purchasing a large size car may signal higher wealth (Skippon and Garwood, 2011). In this context, EV acts as a symbol to construct and communicate a social identity (Graham-Rowe et al., 2012). As Miller (2009) suggests consumers are willing to accept higher prices to communicate such messages to their peers.

*6.4.2. Technological improvements*

Technological aspects of EVs are one of the earliest reported barriers in adopting to EVs from consumers' perspective (Beggs et al., 1981, Hidrue et al., 2011). Low range and high recharge time are found to be the two main shortcomings in EVs compared to the conventional vehicles (Beggs et al., 1981). In this study, range and recharge time are significant in the estimated discrete choice model (refer to section 5.2.1). To evaluate the sensitivity of consumers preference towards these instrumental values, the changes in probability of selecting EV versus improvements in range and recharge time are calculated in scenario 3 and 4 respectively. According to Figure 3, improving range from the base range of 450 km to 700 km (250 km improvement) results in less than 5 percent increase in the probability of selecting EV. Reducing recharge time has even a more trivial impact where reducing recharge time from 75 minutes to 25 minutes (50 minutes reduction) increases the probability of selecting EV by 1 percent.

This observation suggests that the recent improvements in EV industry have alleviated technological concerns around EVs. Although these instrumental values contribute to the utility of purchasing an EV, which means consumers prefer vehicles with longer range and shorter recharge time, range and recharge time don't seem to be barriers towards EV adoption. When Beggs et al. (1981) conducted their study the range for EV was considered between 50 to 100 miles (80.5 to 161 km) and the recharge time between six to eight hours. The limited range and excessive recharge time had put EVs in an inferior position compared to the ICE vehicles with a minimum range of 200 miles (322 km) and a five minute refuel time at the time of their study (Beggs et al., 1981). As another example, when Hidrue et al. (2011) studied individuals willingness to pay for technical improvements in an SP survey, all the levels they defined for vehicle performance was inferior compared to respondents' desired ICE vehicle. However, this unfavourable situation has changed and current commercially available EVs have considerably longer ranges and shorter recharge times. The Tesla Model S, which is the subject vehicle in this study, has a range of 450 km and recharge time of 1.15 hour. Given the long range in this vehicle, consumers can establish recharging facilities at their house, and as the recharging process doesn't need to be supervised, consumers prefer to recharge their vehicles during night times (Skippon and Garwood, 2011).

*6.4.3. Ongoing incentivisation*

Discount on electricity bill is a mechanism for providing ongoing financial incentive to reduce EV's operating costs. The simulation results show an increasing impact on the selecting probability. According to Figure 3, when electricity bill is completely waved, the probability of choosing EV increases by nearly 7 percent for all the target groups. Although positive, the impact of this ongoing incentive is limited. Providing free electricity to EV owner can only increase consumers preference to purchase EV by 7 percent. Previous studies (e.g. see Sovacool and Hirsh, 2009) show that consumers undervalue the lower operating costs of EV and overvalue its high purchase price. Therefore, their calculation of the real price of EV is biased towards higher discount rates and higher payback periods. This behaviour in consumers to overestimate the initial purchase price and under estimate the fuel and maintenance savings from purchasing EVs can be extended to explain lower sensitivity towards ongoing incentives for operating and maintenance costs.

*6.4.4. Market uptake*

In scenario 6 the impact of market uptake on the probability of selecting EV is studied. Although market uptake is not directly controlled or dictated by governments, this exercise helps to put in perspective the effectiveness of other scenarios. The results suggest an increasing impact from market uptake on the probability of selecting EV. The diffusion of innovation (DOI) theory (Rogers, 2010), which is a well-established theory in explaining the spread of ideas and new technologies in societies, asserts that probability of adopting to an innovation will increase as the number of peers who have already adopted to this innovation increases (Rogers, 2010). Bass (1969) proposed a proportional relationship between adoption rate and the cumulative number of adoptions beforehand to formulate the postulated relationship in DOI, and his model was tested in numerous applications (Mahajan et al., 1990). Rogers (2010) believes that there is some level of uncertainty associated with new technologies, and the increase in number of peers who have adopted to a technology decreases this uncertainty. The number of adopters in the society conveys a subjective evaluation to peers. Axsen et al. (2009) refer to this phenomenon as "neighbour impact" which states that the desire to adopt to a new technology increases as it becomes more widespread. In addition to learning from peers, Axsen et al. (2009) mention shift in social norms and increased credibility as other justifications for the positive impact of market share. Krupa et al. (2014) show that many of potential consumers of plugged-in hybrid electric vehicles would wait for a certain level of market uptake before adopting this vehicle. As shown in Figure 3, market uptake is found to have a relatively higher impact compared to discount on electricity bill or technological improvements. This finding is consistent with the generalizations in DOI and reveals the importance of social systems and interpersonal networks in the spread of EVs.

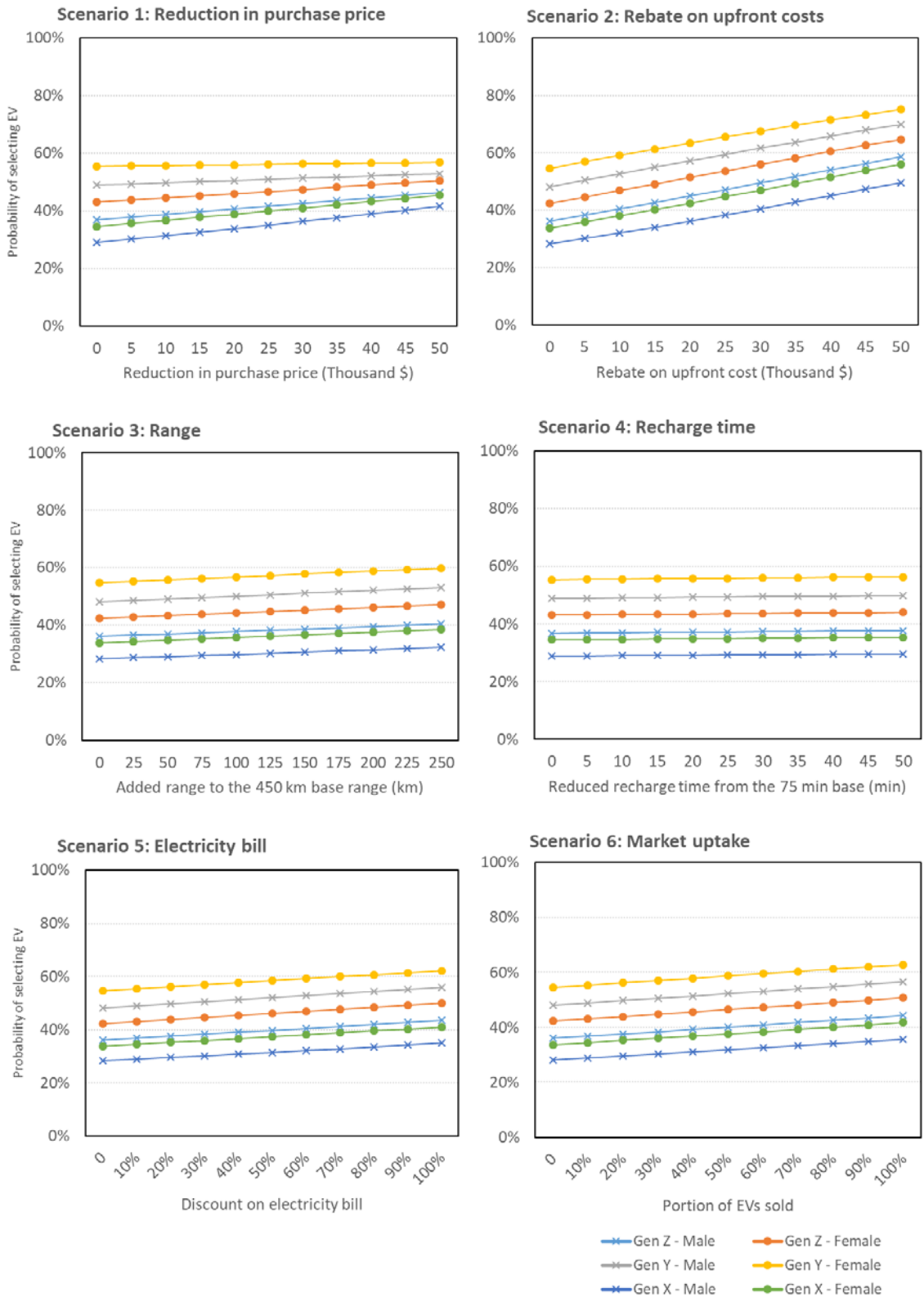

**Figure 4- Sensitivity analysis**

# 7. Conclusion

This paper examined the impact of consumers' *perceived* advantages in EV on their decision to select EVs. Parameters of an integrated choice and latent variable model were estimated using experimental design and stated choice survey from residents of New South Wales Australia. The results show perceived advantage can be represented by three latent variables. The first latent variable, *design*, encompasses consumers' perception from the functional and aesthetic aspects of EVs. Consumers' perception about prestigious style, and efficient and reliable performance are reflected in this *design*. The second latent variable, *environment*, reflects on the perceived environmental impacts of EVs, and the third latent variable, *safety*, expresses the perceived reduction in accident rates by EVs. We found that the perceived advantage for EVs increases by education and employment levels. However, vehicle ownership negatively affects the perceived advantage. Regarding the impact of age, we found a non-linear trend, formulated in quadratic and cubic formats, with a decreasing trend before 50 and an increasing trend after 70.

A logit kernel was estimated as the discrete choice component of ICLV. The results confirmed some findings of earlier studies, such that purchase price, set up cost, operating costs and recharge time have negative affect on the utility of selecting EV, whereas range exert positive impact. The latent variable of *design* is interacted with purchase price found to have a positive coefficient suggesting that those who see higher advantages in EVs' *design,* are willing to pay more for it. *Environment* was interacted with range indicating higher sensitivity towards range for those who see EVs' to be more environmentally friendly. In the same fashion, *Safety* was interacted with large body types suggesting more sensitivity towards body size for those who perceive EVs to be *safer*.

Insights gained from this research shed light on consumers' perception towards EVs, which can assist policy makers to craft more effective strategies for EV widespread. As a showcase, the estimated model was utilised to evaluate the effectiveness of several policies and scenarios. The results showed that technical features of commercially available EVs are no longer a major barrier towards adoption. The range and recharge time for the case study of this paper is set to 450 km and 75 minutes respectively; and consumers' sensitivity to improvements in these features was not considerable. The probability of selecting EV was found more sensitive to financial incentives. Confirming findings in prior studies, consumers value discounts on the purchase price more than discounts on operating costs. Perhaps, the most interesting finding was the dissimilar responses to the same amount of financial incentive, depending on imposing mechanism. Providing incentive to consumers as rebate on purchase price and providing incentive to manufactories to reduce the purchase price result in the same market price for the vehicle; however, consumers associate greater values with the former mechanism.

**Acknowledgement**
This research was funded by